\documentclass[lettersize,journal]{IEEEtran}
\usepackage{amsmath,amsfonts,amsbsy}
\usepackage{algorithmic}
\usepackage{algorithm}
\newtheorem{assumption}{Assumption}
\usepackage{array}
\usepackage[caption=false,font=normalsize,labelfont=sf,textfont=sf]{subfig}
\usepackage{textcomp}
\usepackage{stfloats}
\usepackage{url}
\usepackage{verbatim}
\usepackage{graphicx}
\usepackage{cite}
\begin{document}

\title{Quantifying the Influence of User Behaviors on the Dissemination of Fake News on Twitter with Multivariate Hawkes Processes}

\author{Yichen Jiang and Michael D. Porter

\thanks{Yichen Jiang is with the Department of Systems and Information Engineering, University of Virginia, Charlottesville, 22903, USA (e-mail: yj3us@virginia.edu)}
\thanks{Michael D. Porter is with the Department of Systems and Information Egineering and School of Data Science, University of Virginia, Charlottesville, 22903, USA (e-mail: mdp2u@virginia.edu)}
}

\markboth{}
{Shell \MakeLowercase{\textit{et al.}}: A Sample Article Using IEEEtran.cls for IEEE Journals}


\maketitle

\begin{abstract}
Fake news has emerged as a pervasive problem within Online Social Networks, leading to a surge of research interest in this area. Understanding the dissemination mechanisms of fake news is crucial in comprehending the propagation of disinformation/misinformation and its impact on users in Online Social Networks. This knowledge can facilitate the development of interventions to curtail the spread of false information and inform affected users to remain vigilant against fraudulent/malicious content.
In this paper, we specifically target the Twitter platform and propose a Multivariate Hawkes Point Processes model that incorporates essential factors such as user networks, response tweet types, and user stances as model parameters. Our objective is to investigate and quantify their influence on the dissemination process of fake news. We derive parameter estimation expressions using an Expectation Maximization algorithm and validate them on a simulated dataset. Furthermore, we conduct a case study using a real dataset of fake news collected from Twitter to explore the impact of user stances and tweet types on dissemination patterns. This analysis provides valuable insights into how users are influenced by or influence the dissemination process of disinformation/misinformation, and demonstrates how our model can aid in intervening in this process.
\end{abstract}

\begin{IEEEkeywords}
Hawkes point processes, fake news, online social networks, Twitter, disinformation, dissemination.
\end{IEEEkeywords}

\section{Introduction}\label{intro}
In the digital age, the prevalence of fake news on online platforms has become a pressing concern. Various terms, such as fake news, rumors, and information hoaxes, are used to describe different types of false information, leading to diverse definitions proposed by researchers. Although variations exist, these definitions share fundamental similarities with minor distinctions. Drawing on recent studies \cite{ref1}, we define fake news as “News articles that are intentionally and verifiably false, and could mislead readers". 

The proliferation of fake news topics across social media platforms invariably triggers heated debates among users who hold contrasting perspectives. These discussions possess the potential to incite social events and, in extreme cases, even violence, thereby exerting profound and extensive ramifications on society. 
Generally, a piece of fake news is generated by fake news websites or blogs and shared to social media platforms such as Twitter and Facebook by website users, and spread by platform users. In one example, a piece of fake news on micro-blogging websites was spread within hours and led to mass panic and confusion \cite{ref2}. 
Based on the presence of opinion leaders and trendsetters in social media who possess significant influence over public opinion, it is reasonable to infer that a portion of online users will be swayed by this influence and subsequently propagate it within the online community. However, there are user groups that resist the prevailing voices and are less susceptible to external influence. The diverse opinions held by users in response to fake news stories give rise to distinct attitudes, which we refer to as \textit{user stances}.

The introduction of fact-check articles on social media platforms can have varying effects on user perspectives and the trajectory of discussions. It has the potential to either abruptly terminate the conversation by aligning with the majority view or spark a broader and more extensive discourse if it contradicts prevailing beliefs. The collision of diverse stances and viewpoints within online communities can generate novel modes of interaction and influence among users. However, due to the vast number of participants in complex user networks, comprehending the dissemination process becomes challenging.
Nevertheless, studying the modeling of information cascades, which depict the progression of events in online discussions, is crucial. Such analysis enables us to understand the evolving trends of events and predict their impact, including event popularity. Examining user feedback and interaction provides valuable insights into behavioral and influence patterns, facilitating effective user management on platforms and enhancing user experiences. Particularly concerning the dissemination of fake news, studying the behavioral modes and influence of online users allows for the identification and monitoring of opinion leaders and trendsetters, thereby enabling restrictions on and supervision of malicious users. Additionally, targeted interventions can be provided to misinformed users, contributing to the promotion of social stability. Hence, studying the information cascade on Online Social Networks (OSNs) and the dynamics of user interactions is not only valuable but also essential in understanding and improving the functioning of online platforms, benefiting both users and society as a whole.

This research aims to delve into the intricate causal relationship between user stances, responses concerning the authenticity of fake news, and the subsequent dissemination process on Twitter. 
As previously defined, user stances encapsulate the diverse attitudes users hold towards disseminated fake news stories, discernible from the textual content of their tweets associated with the fake news. Users can respond to fake news on Twitter through various means. Users may: choose to remain passive without responding, initiate conversations by replying to the tweet, or amplify the information within their user networks through retweeting or quoting.
Within a given user network, the behavioral choices associated with different response types (such as retweets or replies) play an important role in the dissemination process, influencing the recipients and the magnitude of the impact. Thus, the type of tweet response to fake news is important and emerges as a significant factor in this study. Specifically, the research will address the following key research questions pertaining to the dissemination of fake news on Twitter:
\begin{itemize}
    \item[1)]Does a user's stance toward the veracity of a fake news article affect the dissemination process? If so, how does the stance of a user impact users with different stances? Are there discernible interaction patterns between users of different stances?   
    \item[2)]To what extent does the dissemination of fake news depend on the type of tweet response? Which particular types of tweets wield the greatest influence in shaping the dissemination process?
\end{itemize}
By exploring these research questions, we aim to unravel the intricate dynamics of fake news dissemination on Twitter, shedding light on the interplay between user stances, responses, and the resulting dissemination patterns. The insights gained from this study will not only enhance our understanding of the complexities surrounding the spread of fake news but also provide valuable guidance for devising effective interventions to mitigate its impact.

We present a novel approach to address the aforementioned challenges by proposing a Multivariate Hawkes Point Processes model (MHP). This model incorporates crucial parameters such as user stances and tweet types, along with the user networks, to accurately capture the dynamics of the fake news dissemination process in online social networks. 
To our best knowledge, this is the first research where user stances and tweet types are considered within the framework of the Hawkes Process model for analyzing fake news dissemination.

Our approach involves introducing parameters associated with user stances and tweet types into the model to examine their influence on the dissemination process of fake news. To quantitatively measure this influence, we develop an Maximum Likelihood Estimation (MLE) procedure using Expectation Maximization (EM) to estimate parameters. This enables us to derive mathematical expressions for each parameter. We validate the effectiveness of our estimation approach by conducting experiments on a realistic simulated dataset.

We collect data from a real fake news event on Twitter and use our fitted model to gain insights into the influences of user stances and tweet types on the dissemination process.  By quantifying the influence of these parameters, we enhance our ability to comprehend and mitigate the dissemination of fake news on online social networks.

The contributions of this article are summarized as follows:
\begin{itemize}
    \item 
    We introduce a novel Multivariate Hawkes Processes model that incorporates user stances and tweet types, along with user networks, as essential parameters. By considering these social dimensions, our model provides a more comprehensive framework for modeling the dissemination process of fake news on Twitter.
    \item 
    We develop parameter estimation procedures for the proposed model using an Expectation Maximization (EM) algorithm. These procedures are validated with a realistic simulated dataset, where the estimated parameters align with the pre-determined values.
    \item
    We evaluate the proposed model and the verified parameter estimation approach using a real dataset of fake news collected from Twitter. This evaluation enables an improved understanding of the influence patterns between users during the dissemination of fake news.
\end{itemize}

 The paper is organized as follows: prior studies in the related area will be reviewed in section \ref{lit}. 
 Modeling procedures with parameter estimation will be introduced in section \ref{method}. A simulated dataset and a real Twitter dataset applied in this paper will be introduced in section \ref{data}, followed by the parameter estimation result and findings demonstrated in section \ref{res}. Limitations and applications will be discussed in section \ref{dis}.

\section{Related Work}\label{lit}
\subsection{Fake News Dissemination Modeling}\label{lit_model}

Fake news dissemination has emerged as a prominent area of research, distinguished by its unique characteristics compared to other types of information diffusion processes. It underscores the significance of user behaviors and user stances within social networks, where users' stances and interactions have a profound impact on the participants, the overall trend, and the consequences of fake news dissemination. For instance, one study utilized propensity scoring to model the dissemination process of fake news, aiming to understand users' sharing behaviors and investigate the causal relationship between user attributes and their susceptibility to fake news \cite{ref3}. Shrivastava \textit{et al.}\cite{ref4} introduced the SIVR (susceptible, infectious, verified, and recovered) model, which captures the dynamics of rumor dissemination among different groups, considering the influence of various measures to counter misinformation. Furthermore, in \cite{ref5}, two models were developed to address malicious behaviors during the fake news dissemination process, accounting for users' opinions towards the information and incorporating the knowledge of its authenticity and trustworthiness as important factors.

Fake news dissemination has also been analyzed with network models.
\cite{ref6} adopted a network simulation model to investigate the possible relationship between echo chamber effects (people prefer to follow like-minded people) and the viral spread of misinformation. They discovered a synergistic relationship between opinions and network polarization on the virality of misinformation. 
Using k-core decomposition, diffusion networks were constructed using tweets gathered prior to the 2016 US Presidential Election \cite{ref7}. The analysis revealed that social bots predominantly dominated the network's core, while fact-checking activities were notably scarce.
A collective influence algorithm in directed networks was developed to uncover how fake news influenced the 2016 US Presidential Election, which found that top influencers spreading traditional center and left-leaning news largely influenced the activities of Clinton supporters while Trump supporters influenced the dynamics of top fake news spreaders \cite{ref8}.
A network approach was used to model fake news spread on Twitter and Weibo (a Chinese micro-blogging website) and found a distinction between the spreading behavior of fake and real news in the early stages of dissemination  \cite{ref9}. The features they propose can be useful for the early detection of fake news. 

These studies have proposed diverse approaches to model the dissemination of fake news on OSNs and answer specific research questions. However, these models have typically focused on specific aspects of the fake news dissemination process, such as user behaviors, opinions, stances, or user networks, while neglecting how they work together to drive or prohibit the spread of fake news.
Furthermore, since information dissemination is a dynamic process unfolding over time, a temporal model would provide a more accurate description of how fake news spreads among online users. Therefore, in this research, we aim to overcome these limitations by employing a Multivariate Hawkes Point Processes, a stochastic temporal model that incorporates all the aforementioned factors, enabling us to more accurately capture and analyze the intricate dynamics of the fake news dissemination process on OSNs.

\subsection{Hawkes Point Processes Modeling on Information Dissemination Process}\label{lit_hawkes}
A Hawkes Point Processes model \cite{ref10} is a statistical model that has been frequently implemented in modeling the information dissemination processes on social media. Specifically they have been used to model user behavior and the overall dissemination process.

\subsubsection{User Behavior}\label{lit_hawkes_ub}
As we discussed in section \ref{intro}, user behavior plays a crucial role in modeling the information dissemination process, regardless of the type of information being propagated. Since online social networks are built for user interactions around emerging topics, it is important to consider user responses and interactions as cognitive behaviors to understand the causal relationship between user behaviors, dissemination mechanisms, and their consequences.

In order to capture the real nature of the information dissemination process on online platforms, it is necessary to incorporate user behaviors, including their responses and interactions, into the modeling framework. 
A Hawkes point process model was constructed in \cite{ref11} to capture the influence between users and account for the influence between multiple types of topics.
Hawkes models have also been applied to trend detection tasks in social networks \cite{ref12} and extended to multiple social networks, considering user-user, topic-topic, and user-topic interactions \cite{ref13}.
Furthermore, a co-evolutionary latent feature process have been developed to accurately capture the evolving nature of user and item features in recommendation systems for online service websites, explicitly incorporating user-item interactions \cite{ref14}. A Fourier-based Multidimensional Hawkes Process has been used to investigate correlations between online users' activities and evaluated for activity prediction on platforms like Github and Metafilter \cite{ref15}. Additionally, Hawkes models that incorporate user profile features and networks have been applied to Twitter data. 

These studies highlight the versatility and applicability of Hawkes Processes for modeling user behavior, interactions, and networks on OSNs.
However, it is important to note that user behavior encompasses a broad range of activities related to information dissemination, which can vary across different social media platforms and types of information being propagated. This observation has motivated our research to focus on understanding user activities specifically during the dissemination of disinformation on Twitter, with the aim of providing a more precise characterization of the process.

\subsubsection{Dissemination of Disinformation}\label{lit_hawkes_dd}
Hawkes Point Process models have also been applied in modeling the overall dissemination of disinformation effectively. A two-stage Time-Dependent Hawkes Process (TiDeH) model was built for characterizing the process of fake news dissemination on Twitter before/after fact-checking occurs \cite{ref16}, which considered tweet posts as the user behaviors over time. A related study applied a Multivariate Hawkes Process incorporating user networks to measures the influence rate between online users to model the process of rumor propagation on Twitter \cite{ref17}. A classification model was developed for user stances of rumors on Twitter using Hawkes Processes.
Similarly, Multivariate Hawkes models using textual-based base intensity was built for rumor stance classification \cite{ref19}. User behaviors of pathogenic user accounts on social media have been studied with respect to their posted URLs compared with normal accounts to analyze the subsequent negative influence using the Hawkes models \cite{ref20}. A hybrid model combining Hawkes and and topic modeling was developed that incorporates temporal and textual features of tweets such as the number of retweets per user for detecting fake retweeters \cite{ref21}. To combat fake news, the Hawkes models was also adapted for a reinforcement learning approach to detecting fake news and making interventions \cite{ref22,ref23} where \cite{ref22} considers the user networks and the user behaviors of posting fake news and mitigation events, and \cite{ref23} developed several models with the influence of past events, the mutual influence between users, and the possible political bias. \cite{ref25} proposes a simulation approach to generate tweet data associated with fake news spreading which considers the user stances, networks, and different tweet types.

The preceding section provided a summary of the existing studies focusing on the dissemination of disinformation, specifically in the context of Twitter and other social media platforms. 
These studies have primarily examined the dissemination process in relation to fake news and rumors using variations of a Hawkes point process model.
In OSNs, the user network is the framework that enables the both unidirectional and bidirectional influence among users.
These connections play a crucial role in determining which information is propagated, who is impacted by it, and when such influence occurs. Additionally, Twitter users have various ways to respond to fake news stories, including retweeting, quoting, or replying to the original fake news tweet while expressing their own stances towards the content.
To effectively address our research questions, we have opted to model the dissemination process of fake news on Twitter by incorporating user networks, user tweet types, and user stances towards the fake news stories as the essential model parameters of a Hawkes Point Process model. Further details regarding the intricacies of our model will be elaborated upon in the subsequent section.

\section{Methodology}\label{method}
This section will introduce the Multivariate Hawkes Point Processes model that applies to the modeling of fake news dissemination on Twitter. 
Maximum Likelihood Estimation (MLE) using an Expectation Maximization (EM) algorithm is applied to derive the expression of estimating the model parameters which provides the understanding of the mechanism of the fake news dissemination process on Twitter.

\subsection{Modeling}\label{method_model}
This paper proposes a Multivariate Hawkes Point Processes model based on the simulation work in \cite{ref25} with appropriate adaptation on the model expression for the derivation of parameter expression using MLE with EM algorithm to estimate the parameter values.

\subsubsection{Hawkes Point Processes}\label{method_model_hawkes}
The Hawkes Point Processes model carries the assumption that subsequent events in the process are generated under the influence of prior events.
These models have been frequently applied in the social media domain to describe the information dissemination process and information cascades. These models of the intensity, or event rate, take the following standard form:
\begin{equation}
    \lambda(t) = \lambda_0(t)+\sum_{i:t_i<t}\alpha g(t-t_i)
\end{equation}
The model intensity $\lambda(t)$ is composed of the immigrant process and the self-exciting process. The immigrant rate function $\lambda_0(t)$ determines the immigrant process over time in which immigrants are the initial fake news events that stimulate the process, and $\sum_{i:t_i<t}\alpha g(t-t_i)$ describes the self-exciting process that stimulates the occurrence of subsequent, or descent, events. 
The descendant event rate, $\alpha$ is the branching factor that measures the expected number of new occurrences excited by any given occurrences, and $g(t-t_i)$ refers to the kernel function which determines the length of time a prior event can influence the future event rate.

\subsubsection{Multivariate Hawkes Processes Modeling on Twitter Fake News Dissemination Process}\label{method_model_mul-hawkes}
Based on the standard form of Hawkes Point Processes, the Multivariate Hawkes Processes model describes multiple Hawkes processes occurring simultaneously, with possible interaction and influence between processes. 
Specifically, this paper adapts the Multivariate Hawkes Processes model to the fake news dissemination process on Twitter by incorporating the user networks, user stances toward the fake news stories, and tweet types as the model parameters co-acting on the process.
Each tweet, regardless of the tweet type, will be considered an event that may encourage subsequent tweets to be generated.

As discussed in the previous sections, we are most interested in learning how the type of tweet and the user stance influence fake news dissemination.
Twitter allows four different tweet types: original tweets, retweets, quotes, and replies. Original tweets are the initial tweets with disinformation. The disinformation may originate from Twitter or outside the Twitter platform but is delivered to Twitter by an initiator. Retweets, quotes, and replies are the subsequent tweets generated from the original tweets or other subsequent tweets, where retweets are the retweeted tweets without users' comments, quotes are retweeted tweets with comments from users, and replies are the responding tweets towards a prior tweet with user comments. Original tweets correspond to the immigrants in Hawkes Process theory, and the other three tweet types are modeled as descendants in the model. Since we want to explore the influence between different stances of tweets during the process of fake news dissemination, the Multivariate Hawkes Processes model allows us to characterize the dissemination process of each type of user stance towards the fake news story with the interplay between stances. 

\begin{table}[!t]
\caption{Parameter Notation and Definition\label{tab:param_def}}
\centering
\begin{tabular}{|c|c|}
\hline
\textbf{Notation}&\textbf{Definition}\\
\hline
$j$ & Tweet number of a prior tweet (event)\\
\hline
$i$ & The potential user\\
&that has been influenced by the prior tweets\\
\hline
$u_j$ & The user of event $j$\\
\hline
$r\in R$ & Tweet types\\
& $R = \{ \text{original tweets, retweets, quotes, and replies} \}$ \\
\hline
$k\in K$ & User stances\\
&  $K = \{ \text{supporting, not-supporting stances}\} $\\
\hline
$t$ & The current time\\
\hline
$T$ & Total time the dissemination process is observed\\
\hline
$Z(t)$ & The aggregation of tweets generated before time $t$\\
\hline
$\mu_k$ & The basic immigrant rate for users with stance $k$\\
\hline
$x$ & Scale parameter of the Truncated Exponential Distribution\\
\hline
$\delta_{r_j}$ & The influence factor of the tweet type of the prior tweet $j$\\
\hline
$\beta_{i,u_j}$ & The factor of user relationships between user $i$ and user $u_j$\\
\hline
$\gamma_{k_j,k}$ & The influence factor between stance $k$ \\
&and stance $k_j$ (stance of event $j$)\\
\hline
$p_r$ & The probability that\\
&user $i$ will generate a tweet of tweet type $r$\\
\hline
$g_k(t-t_j)$ & The kernel function\\
\hline
$\omega_k$ & The decay parameter of the exponential kernel function\\
\hline
$\lambda(t)$ & Process intensity at current time $t$\\
\hline
$L(t;\theta)$ & Likelihood function\\
\hline
$Q(T;\theta)$ & The expectation of the log-likelihood function\\
\hline
$p_{jj}$ & The probabilities that event $j$ is an immigrant\\
\hline
$p_{jl}$ & The probabilities that event $l$ is a descendant that\\
&  influenced by any prior tweet $j$\\
\hline
\end{tabular}
\end{table}

The definition of parameters with their notation are presented in Table \ref{tab:param_def}. Consider the case that user $i$ generates a tweet with stance $k$ of type $r$ where $r=\mathrm{original\ tweet}$ (immigrants), the corresponding event intensity is calculated as follows:
\begin{equation}
    \lambda_{kri}(t-t_j;\theta,Z(t))=\frac{\mu_k(t)}{U}
\end{equation}
where $U$ is the total number of users in the social networks who engage in the fake news dissemination process, $Z(t)$ is the aggregation of the tweets that were generated before current time $t$, $t_j$ is the arrival time of the tweet $j$, $k$ is the user stance towards a fake news story which will be illustrated later in detail, and $\mu_k(t)$ is the immigrant function that controls the rate of generating immigrants, which are the original tweets in the Twitter scenario, with the following form:
\begin{equation}
\mu_k(t) = \mu_k\frac{xe^{-xt}}{1-e^{-xT}}
\end{equation}
It models the process of generating original tweets through a base rate $\mu_k$ and a truncated exponential distribution with scale parameter $x,\ 0<x<T$, and bounded between 0 and total observation time $T$. Particularly, the truncated exponential distribution is applied to imitate the arrival time of the original tweets with a right-skewed pattern based on our observation from a real Twitter fake news dataset which will be introduced in section \ref{data}.

The intensity function for user $i$ generating a tweet with stance $k$ and type $r$ where $r$ is a retweet, quote, or reply can be expressed:
\begin{equation}
\lambda_{kri}(t-t_j;\theta,Z(t))=\sum_{j=1}^{N(t)}\delta_{r_j}\beta_{i,u_j}\gamma_{k_j,k}p_rg_k(t-t_j)\label{eq:intensity_sub}
\end{equation}
We model this self-exciting process of generating descendants with the influences of tweet type $\delta_{r_j}$ of prior tweet $j$, user relationship $\beta_{i,u_j}$, stance influence $\gamma_{k_j,k}$ from prior tweet $j$, probability $p_r$ of generating a new tweet in tweet type $r$, and the kernel function determines the decay of the influence of tweet $j$ over time. Specifically, for the user stances $k\in K$, this paper considers two possible stances during the dissemination process: a) supporting stance: the user supports the tweet; 
b) not-supporting stance: the user does not support the tweet. This can include denying, questioning, or commenting (no stance toward the fake news story).

For the factor of user relationship $\beta_{i, u_j}$, we have:
\begin{equation}
\beta_{i, u_j} = 
\begin{cases}0.95\quad\ \text{if user } $i$ \text{ follows the user of event } $j$ \\
 0.05\quad \  \text{user } $i$\ \text{is able to see the user of replies}\\
\quad\quad\quad \ \text{(event) }$j$ \text{ under the tweets, including}\\
\quad\quad\quad \ \text{original tweets, quotes, and retweets from}\\
\quad\quad\quad \ \text{someone user } $i$ \text{ follows}  \\
 0\quad\quad\ \ \text{otherwise}
\end{cases}
\end{equation}
\begin{equation}\label{eq:p_r}
p_{ki}(r)=p_r=\begin{cases}p_1\quad \  \text{if the new tweet is a retweet}\\ 
p_2\quad\ \text{if the new tweet is a quote}\\  
1-p_1-p_2\ \ \text{if the new tweet is a reply}\\ 
0\quad\quad \text{if the new tweet is an original tweet}
\end{cases}
\end{equation}
such that
\begin{equation}
\sum_{r\in R}p_{ki}(r)=\sum_{r\in R}p_r=p_{ori}+p_{ret}+p_{quo}+p_{rply} = 1
\end{equation}
This indicates that the probability of generating any tweet type follows the distribution in equation \ref{eq:p_r}, and the probability of generating an original tweet given prior tweets is 0, which means the emergence of original tweets is not influenced by the self-exciting process.
We further simplify the sum of the user relationship as:
\begin{equation}
\sum_{i}^{U}\beta_{i,u_j} = n_j
\end{equation}
where $n_j$ is a function that combines the values of user influences between the current user $i$ and each of the users $u_j$ who generated the prior tweet $j$. Note that $n_j$ is just a parameter for purposes of notation simplification, and we still need to calculate the intensity of a prior event by multiplying the user relationship value $\beta_{i,u_j}$ by other parameters according to Equation~\ref{eq:intensity_sub}.

Considering that the tweets in different stances are disseminated simultaneously, we can model the overall process by modeling each specific stance $k$ with the interplay between stances, and aggregate the intensities of all the stances to the overall intensity:
\begin{align}
\lambda_{k}(t;\theta) &=\sum_i^U\left(\frac{\mu_{k}(t)}{U}+\sum_{r\in R}\sum_{j=1}^{N(t)}\lambda_{kri}(t-t_j;\theta,Z(t))\right)\\
=\mu_k&\frac{xe^{-xt}}{1-e^{-xT}}+\sum_{j=1}^{N(t)}\delta_{r_j}\gamma_{k_j,k}g_k(t-t_j)\sum_i^U\beta_{i,u_j}\sum_{r\in R}p_r\\
=\mu_k&\frac{xe^{-xt}}{1-e^{-xT}}+\sum_{j=1}^{N(t)}\delta_{r_j}\gamma_{k_j,k}n_jg_k(t-t_j)
\end{align}
such that the expression of the overall intensity at time $t$ can be derived as follows:
\begin{equation}\label{lambda}
\lambda(t;\theta)= \sum_{k\in K}\left(\mu_k\frac{xe^{-xt}}{1-e^{-xT}}+\sum_{j=1}^{N(t)}\delta_{r_j}\gamma_{k_j,k}n_jg_k(t-t_j)\right)
\end{equation}
and is used for calculating the likelihood.

\subsection{Parameter Estimation}\label{method_est}
In order to learn about the dissemination process for a particular fake news event, the model needs to fit the event data. 

We derive a Maximum Likelihood Estimation procedure using an Expectation Maximization approach that recursively optimizes the expectation of the log-likelihood function and updates the parameters. 

\subsubsection{Model Likelihood}\label{method_est_mle}
The likelihood Function for the overall intensity of the Hawkes process:
\begin{equation}
    L(t;\theta) = \prod_{j=1}^{N(t)}\lambda(t_j;\theta)e^{-\int_0^T\lambda(t_j;\theta)\mathrm{d}t}
\end{equation}
such that the corresponding log-likelihood function can be expressed as follow:
\begin{align}
&\operatorname{log}L(t;\theta) = -\int_0^T\lambda(t;\theta)\mathrm{d}t+\sum_{j=1}^{N(t)}\operatorname{log}\lambda(t;\theta)\\
&\begin{aligned}
=&\sum_{k\in K}\Bigg[-\mu_k-\sum_{j=1}^{N(t)}\int_{t_j}^{T}\delta_{r_j}\gamma_{k_j,k}n_jg_k(t-t_j)\mathrm{d}t\\
&+\sum_{j=1}^{N(t)}\operatorname{log}\left(\mu_k\frac{xe^{-xt}}{1-e^{-xT}}+\sum_{l=1}^{j-1}\delta_{r_l}\gamma_{k_l,k_j}n_lg_k(t_j-t_l)\right)\Bigg]
\end{aligned}
\end{align}

\subsubsection{Expectation Maximization}\label{method_est_em}
The Expectation Maximization algorithm (EM) \cite{ref26} is a classical approach for parameter estimation for Hawkes models which finds the maximum likelihood under the case of unobserved data and latent variables from the dataset. EM algorithm alternates between the expectation step (E-step) and the maximization step (M-step), where the E-step calculates the expectation of the log-likelihood function which is also called the $Q$ function, while the M-step estimates the parameters by maximizing the $Q$ function. This repeats until the $Q$ function doesn't significantly change.

The $Q$ function is transformed from the log-likelihood function through Jensen's inequality, which takes the following form:
\begin{equation}
\begin{split}
Q(T;\theta) =&\sum_{k\in K}\sum_{j=1}^{N(t)}p_{jj}^{k}\operatorname{log}\left(\mu_k\frac{xe^{-xt}}{1-e^{-xT}}\right)-\sum_{k\in K}\mu_k\\
&+\sum_{k\in K}\sum_{j=1}^{N(t)}\sum_{l=1}^{N(t_j)}p_{jl}^{k}\operatorname{log}\left[\delta_{r_j}\gamma_{k_j,k}n_jg_k(t-t_j)\right]\\
&-\sum_{k\in K}\sum_{j=1}^{N(t)}\int_{t_j}^T\delta_{r_j}\gamma_{k_j,k}n_jg_k(t-t_j)\mathrm{d}t+C
\end{split}
\end{equation}
where $p_{jj}$ and $p_{jl}$ are the probabilities that event $j$ is an immigrant (original tweet), or it is a descendant (retweets, quotes, or replies) caused by any prior influential tweets respectively. More details of the derivation of the $Q$ function are presented in Appendix \ref{app:mle}.

The components of the EM algorithm are estimated as follows:
\begin{enumerate}
\item [a)] Expectation Step\\
The E-step calculates the $p_{jj}$ and $p_{jl}$. Using an exponential decay function with parameter $\omega_k$ gives:
\begin{align}
&{p_{jj}^k}^{(s+1)}=\nonumber \\
&\frac{\mu_k^{(s)}\frac{xe^{-xt_j}}{1-e^{-xT}}}{\mu_k^{(s)}\frac{xe^{-xt_j}}{1-e^{-xT}}+\sum_{l=1}^{j-1}\delta_{r_l}^{(s)}{\gamma_{k_l,k_j}^{(s)}}n_l\omega_k^{(s)}e^{-\omega_k^{(s)}(t_j-t_l)}}\\
&{p_{jl}^{k}}^{(s+1)}=\nonumber\\
&\frac{\delta_{r_l}^{(s)}{\gamma_{k_l,k_j}^{(s)}}n_l\omega_ke^{-\omega_k(t_j-t_l)}}{\mu_k^{(s)}\frac{xe^{-xt_j}}{1-e^{-xT}}+\sum_{l=1}^{j-1}\delta_{r_l}^{(s)}{\gamma_{k_l,k_j}^{(s)}}n_l\omega_k^{(s)}e^{-\omega_k^{(s)}(t_j-t_l)}}
\end{align}
\item [b)] Maximization Step\\
The $Q$ function represents the lower bound of the log-likelihood function, and the M-step estimates the parameters to maximize this lower bound.

The maximum value of parameter $\theta$ occurs when $\frac{\partial Q(T;\theta)}{\partial\theta}=0$ such that we can derive the estimation expression for the parameters:
\begin{equation}
    \mu_k^{(s+1)}=\frac{\sum_{j=1}^{N(t)}{p_{jj}^{k}}^{(s)}}{T}
\end{equation}
\begin{equation}
    {\gamma_{k',k}}^{(s+1)}=\frac{\sum_{j:r_j\in\{quo,rply\}}^{N(t)}\sum_{l:k_l=k'}^{j-1}{p_{jl}^k}^{(s)}}{\sum_{k\in K}\sum_{j:r_j\in\{quo,rply\}}^{N(t)}\sum_{l:k_l=k'}^{j-1}{p_{jl}^k}^{(s)}}
\end{equation}
\begin{equation}
    \omega_k^{(s+1)}=\frac{\sum_{j=1}^{N(t)}\sum_{l=1}^{j-1}{p_{jl}^k}^{(s)}}{\sum_{j=1}^{N(t)}\sum_{l=1}^{j-1}{p_{jl}^k}^{(s)}(t_j-t_l)}
\end{equation}
\begin{equation}
    \delta_{r}^{(s+1)}=\frac{\sum_{k\in K}\sum_{j=1}^{N(t)}\sum_{l:r_l=r}^{j-1}{p_{jl}^k}^{(s)}}{\sum_{k\in K}\sum_{j:r_j=r}^{N(t)}\gamma_{k_j,k}n_j}\\
\end{equation}

\item [c)] $Q$ function\\
After the E-step and M-steps at each iteration $s$, the $Q(T;\theta)$ is updated and checked for convergence:
\begin{align}
&Q^{(s+1)}(T;\theta)\approx\sum_{k\in K}\sum_{j=1}^{N(t)}{p_{jj}^k}^{(s)}\operatorname{log}(\mu_k^{(s)}\frac{xe^{-xt}}{1-e^{-xT}})\nonumber \\
&+\sum_{k\in K}\sum_{j=1}^{N(t)}\sum_{l=1}^{j-1}{p_{jl}^k}^{(s)}\operatorname{log}\left(\delta_{r_l}^{(s)}\gamma_{k_l,k_j}^{(s)}n_l\omega_k^{(s)}e^{-\omega_k^{(s)}(t_j-t_l)}\right)\nonumber \\
&-\sum_{k\in K}\sum_{j=1}^{N(t)}\delta_{r_j}^{(s)}\gamma_{k_j,k}^{(s)}n_j-\sum_{k\in K}\mu_k^{(s)}+C
\end{align}
We compare $Q^{(s+1)}$ with $Q^{(s)}$ until 
\begin{equation}
    |Q^{(s+1)}(T;\theta)-Q^{(s)}(T;\theta)| \leq \epsilon
\end{equation}
which indicates that the optimal solutions for the parameters have been reached.\\
\end{enumerate}

\section{Dataset}\label{data}
In this section, we present the dataset that we utilized to validate our developed parameter estimation procedures and examine the patterns and mechanisms of interaction during the dissemination process of fake news on Twitter.

\subsection{Simulation Dataset}\label{data_sim}
\begin{figure}[!t]
\centering
\includegraphics[width=3.5in]{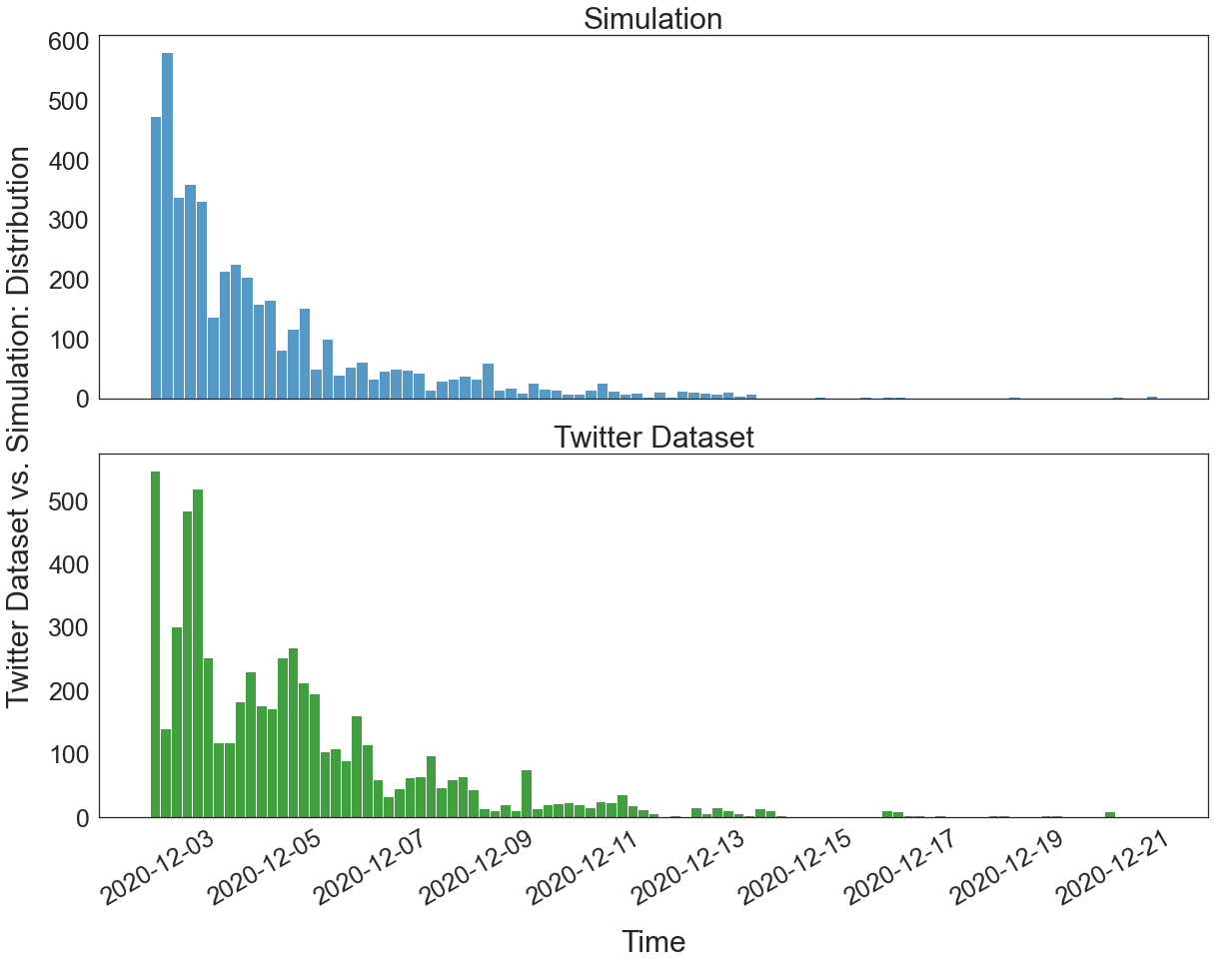}
\caption{Event distribution: Simulation vs. Real Twitter Dataset}
\label{fig:data}
\end{figure}

The simulation dataset applied in this paper is a dataset generated through the simulation of a Hawkes model proposed in \cite{ref26} using the intensity-based approach developed from Ogata's thinning method \cite{ref27} and cluster-based simulation approach \cite{ref28}. This simulation approach considered the tweet types and user stances in the dissemination process of fake news on Twitter in a simulated user network with pre-determined parameter values. The simulated dataset was constructed to mimic the properties of a real fake-news event (detailed in the next section). It contains 3470 simulated events with 2925 supporting tweets, 545 not-supporting tweets, 1011 original tweets, 1527 retweets, 297 quotes, and 344 replies, which have been summarized in the upper half of Table \ref{tab:dataset} with the distribution of the data showing in \ref{fig:data}. 
We use this simulated data, generated by a process with known parameters, to validate our estimation procedure. 

\begin{table}[htbp]
\caption{Event Count}
\begin{center}
\begin{tabular}{|c|c|c|c|c|c|}
\hline
{\textbf{Simulated}}&\multicolumn{5}{|c|}{\textbf{Tweet Types $\boldsymbol{r}$}} \\
\cline{2-6}
{\textbf{Dataset}}&\textbf{Original}&\textbf{Retweet}&\textbf{Quote}&\textbf{Reply}&\textbf{Total}\\
 \hline
\textbf{Supporting}&895&1527&231&272&2925\\
 \hline
\textbf{Not-Supporting}&116&291&66&72&545\\
 \hline
\textbf{Total}&1011&1818&297&344&3470\\
\hline
{\textbf{Real}}&\multicolumn{5}{|c|}{\textbf{Tweet Types $\boldsymbol{r}$}} \\
\cline{2-6}
{\textbf{Dataset}}&\textbf{Original}&\textbf{Retweet}&\textbf{Quote}&\textbf{Reply}&\textbf{Total}\\
 \hline
\textbf{Supporting}&812&3175&59&708&4754\\
 \hline
\textbf{Not-Supporting}&181&720&14&240&1155\\
 \hline
\textbf{Total}&993&3895&73&948&5909\\
\hline
\end{tabular}
\label{tab:dataset}
\end{center}
\end{table}

\subsection{Real Twitter Dataset}\label{data_real} 
\subsubsection{Data Collection}\label{data_real_collect}
A real Twitter dataset collected through the Twitter API v1 and v2 is used as an example of how fake news disseminates over time on Twitter. As summarized in Table \ref{data_real}, this dataset contains 5909 tweets that relate to one of the fake news stories that occurred during December 2020 which claimed that Covid vaccine causes female sterilization and has been debunked later on December 5th, 2020 \cite{ref29}. The dataset covered the data from the start of the process until December 21st, 2020 with 4754 supporting tweets, 1155 not-supporting tweets, 993 original tweets, 3895 retweets, 948 replies, and 73 quotes, which is summarized in the lower half of Table \ref{tab:dataset} with the data distribution showing in \ref{fig:data}. 
We chose this particular fake news story for our case study data collection because its content is highly relevant to users' personal lives and their connections with family and friends. Users are likely to "at" (Twitter function to notify someone of a tweeted tweet) their friends and family members who are connected to them in order to share and review the content. As a result, the dissemination process of this fake news story is strongly influenced by user networks, making it a more accurate representation of the characteristics of disinformation dissemination on online social networks (OSNs). By modeling the fake news dissemination process using Multivariate Hawkes Processes, we aim to provide insights that address our research questions and shed light on the dynamics of information spread in such scenarios.

The user relationships were collected for each user that engaged in the propagation of the above fake news story to form the user network and study the information cascades. Based on the dataset, 4577 users were found that generated the tweets associated with the fake news story, however, only 2963 users' relationships were found and collected through the data collection due to some accounts being suspended at the time of data collection as well as user privacy settings of being able to be accessed through Twitter API. For those tweets in which the user account information is missing, there is no obvious influence direction we could obtain from the user relationships. In such a case, we will rely on the information attached to the tweet of whether it is a retweet/quote/reply of a prior tweet in our dataset, and use this prior tweet information to understand the dissemination process connectively.
If the current tweet is a retweet/quote/reply of a prior tweet that does not belong to our dataset, in other words, an irrelevant tweet towards the fake news story, then we will assume that the user received the information about the fake news story from other sources such as hashtags or searching from keywords, and we will consider all tweets that tweeted prior than the current tweet as the prior influential tweets.

\subsubsection{Data Labelling}\label{data_real_label}
\begin{figure}[!t]
\centering
\includegraphics[width=3.5in]{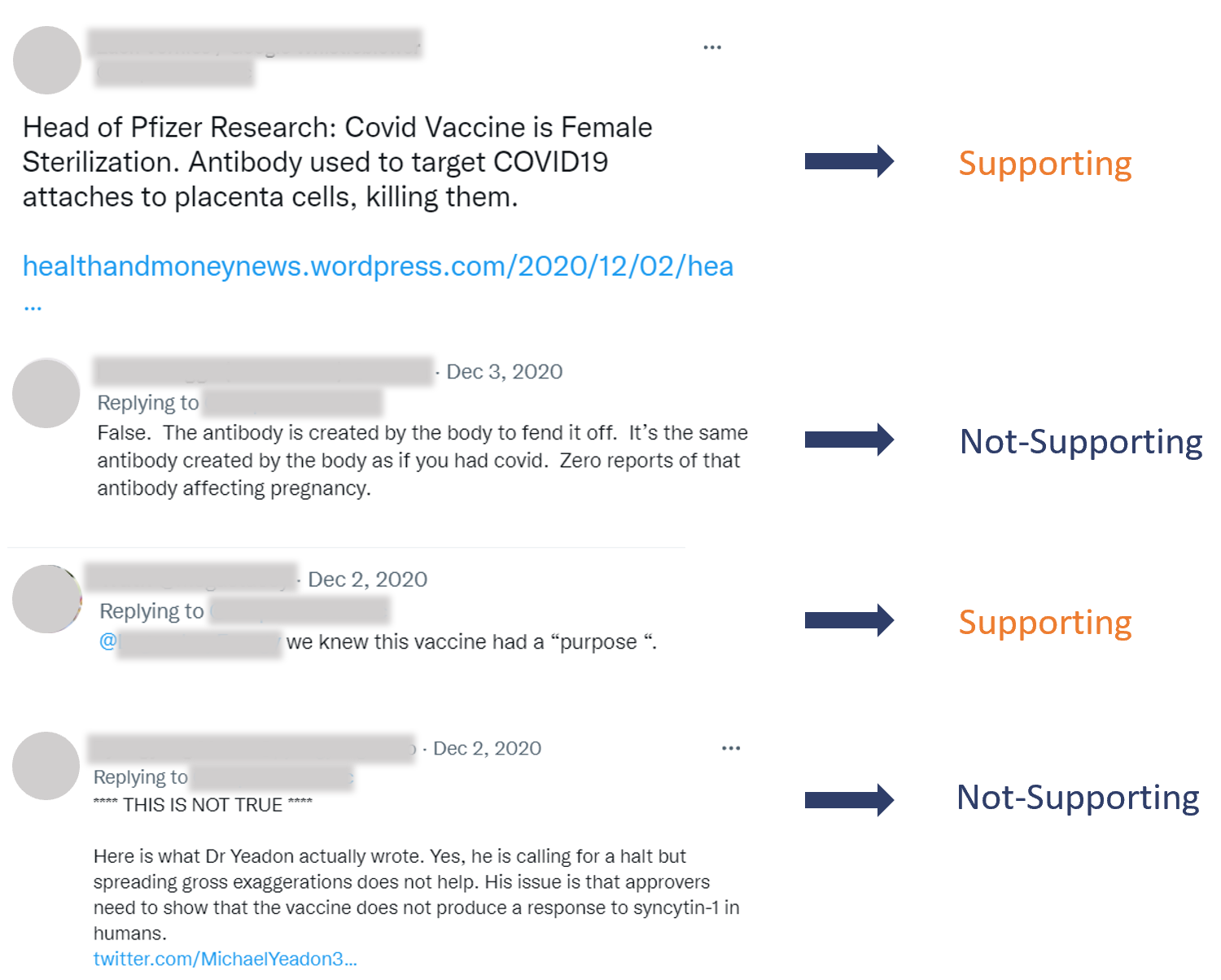}
\caption{An Example of Tweet Stances}
\label{fig:label}
\end{figure}

All the labels of tweet stances are assigned manually. Figure \ref{fig:label} shows an example of our labeling work of tweet stances where an original tweet with the supporting stance was responded by replies tweeted by multiple users with supporting or not-supporting stances. Note that the not-supporting stance defined in this paper represents the attitude of not showing an obvious supporting stance from the tweet content, which includes the case of showing denying stance, questioning the fake news story, and commenting on the fake news without any particular stance. 

Particularly, we hold two assumptions for labeling the user stances toward the fake news story revealed from their tweets. These assumptions are similar to those stated in \cite{ref25} which we state again to help explain our labeling process.
\begin{assumption}\label{asp:ori}
    The tweets of the original sources of the fake news with no additional comments should hold a supporting stance as the sources.
\end{assumption}
\begin{assumption}\label{asp:ret}
The stance of a retweet should be the same as the stance of the event that triggers it.
\end{assumption}
These two assumptions helped us to label the tweet stance based on the tweet cascades.

Two annotators were asked to perform the labeling work on the tweet stance. The first annotator was asked to label the stance of all tweets, and the second annotator checked the labeling results by taking a random sample and labeling 100 tweets. These labels were compared to the labels assigned by the first annotator. Among the 100 samples that were labeled by both annotators only 8 labels had a different stance. 
This indicates a roughly 92\% accuracy on the data labeling work which is acceptable for understanding the main mechanisms driving the dissemination. 

\section{Results}\label{res}
This section presents the results of parameter estimation by applying the EM algorithm on both the simulated dataset and the real Twitter dataset. The code of the EM algorithm derived for the model can be accessed online \cite{ref31}.

\subsection{Parameter Estimation on Simulated Dataset}\label{res_sim}
Table \ref{tab:est_sim} compares the true values and the estimation result for each parameter in the model where the parameters with fixed values are not included in the table. As the table shows, the EM algorithm estimates most of the parameters accurately such as the base immigrant rates for both supporting $\mu_s$ and not-supporting stances $\mu_n$, the scale parameter $x$ of the truncated exponential distribution, and the influence factor between stances where $\gamma_{ns}=0.3473$ and $\gamma_{nn}=0.6527$ slightly deviate from the true values (0.5 and 0.5) due to the very limited number of not-supporting tweets generated in the simulation. The sum of $\gamma_{ns}$ and $\gamma_{nn}$, as well as $\gamma_{ss}$ and $\gamma_{sn}$ equal to 1 which matches the meaning we defined for $\gamma$ as it represents the conditional probabilities of generating supporting and not-supporting tweets given any specific stance should sum up to 1.

The estimation of influence factors of tweet types does not match their true values closely due to the edge effect of setting relatively smaller values for them, and the estimations of decay parameters deviating from the true values may be due to the inaccurate estimations of the influence factors of tweet types. However, the relative relationships between the same parameter type are almost correct which reveals the fact that original tweets take the dominant place in influencing new tweets, followed by retweets and the other two tweet types, and the supporting tweets will hold a longer influence than the not-supporting tweets.
\begin{table}[htbp]
\caption{Parameter Estimation: Simulated Dataset}
\begin{center}
\begin{tabular}{|c|c|c|c|}
\hline
\multicolumn{2}{|c|}{\textbf{Parameter Names}}&\textbf{True Values}&\textbf{Estimation} \\
\hline
\textbf{Immigrant Rate}& $\mu_s$&0.15&0.1492\\
\cline{2-4}
$\boldsymbol{\mu_k}$& $\mu_n$ &0.015&0.0193\\
\hline
\textbf{Truncated} & lower bound&0&/ \\
\cline{2-4} 
\textbf{Exponential}& upper bound&6000&/ \\
\cline{2-4} 
\textbf{Distribution} & scale $x$&1000&969.9030 \\
\hline
& $\delta_{ori}$&$1.5\times10^{-3}$&$4.8856\times10^{-3}$  \\
\cline{2-4} 
\textbf{Influence Factor}&$\delta_{ret}$&$2\times10^{-5}$&$5.1844\times10^{-4}$  \\
\cline{2-4} 
\textbf{of Tweet Type} & $\delta_{quo}$&$2.5\times10^{-6}$&$5.8166\times10^{-35}$  \\
\cline{2-4} 
$\boldsymbol{\delta_r}$& $\delta_{rply}$&$5\times10^{-6}$&$1.0185\times10^{-52}$  \\
\hline
& $\gamma_{ss}$&0.9&0.8841  \\
\cline{2-4} 
\textbf{Influence Factor}& $\gamma_{sn}$&0.1&0.1159  \\
\cline{2-4} 
\textbf{between Stances} & $\gamma_{ns}$&0.5&0.3473  \\
\cline{2-4} 
$\boldsymbol{\gamma_{k',k}}$& $\gamma_{nn}$&0.5&0.6527 \\
\hline
\textbf{Decay Parameter}&$\omega_s$&3&2.0096\\
\cline{2-4}
$\boldsymbol{\omega_k}$& $\omega_n$& 1.5&0.8690\\
\hline
\end{tabular}
\label{tab:est_sim}
\end{center}
\end{table}

\subsection{Parameter Estimation on Real Twitter Dataset}\label{res_real}
We collected user relationships from the follower lists of each user. This information helps us understand which users are likely to be impacted by prior tweets. As previously mentioned, due to privacy settings and blocked accounts, some user followers and following lists may be incomplete or missing.
Furthermore, this directional information also aids in identifying the influence paths between prior tweets and the current tweet, allowing us to track how the dissemination process unfolds over time.
The dataset used in our analysis spans a period of 20 days, equivalent to 480 hours.

\begin{figure*}
  \includegraphics[width=\textwidth,height=7.5cm]{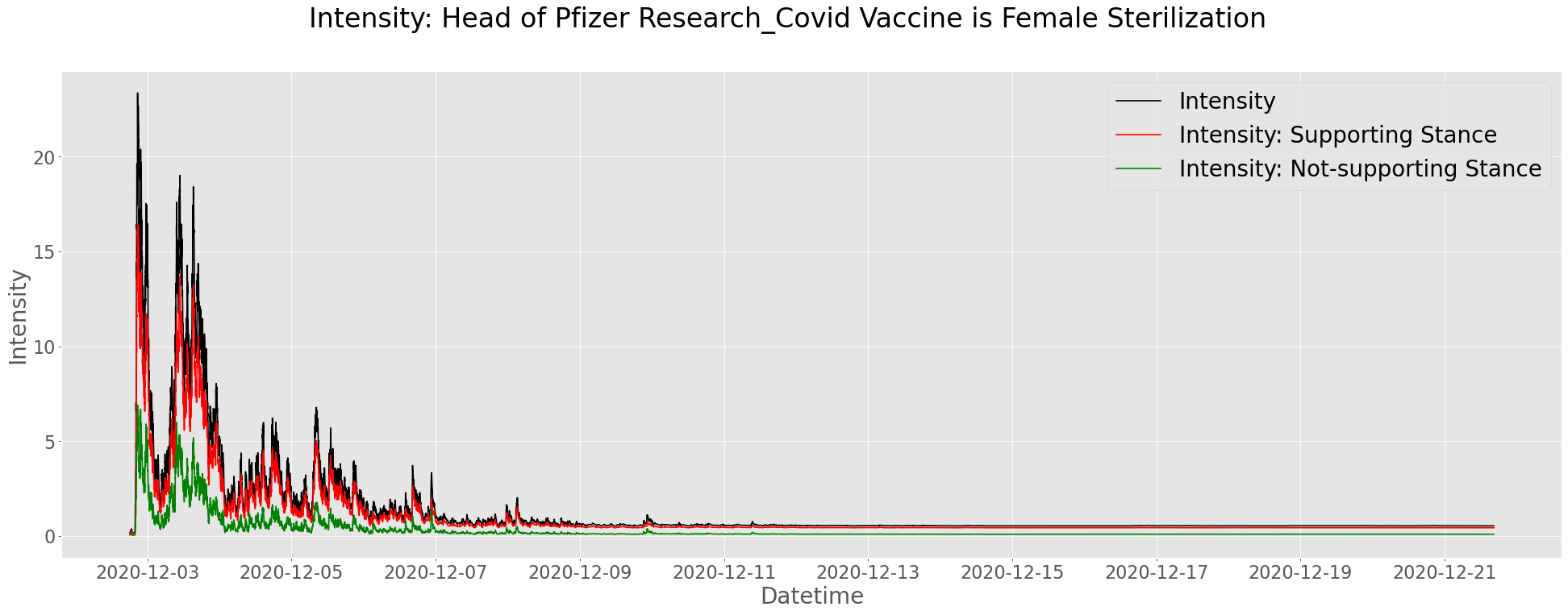}
  \caption{The Intensity of the Fake News Story: 'Head of Pfizer Research: Covid Vaccine is Female Sterilization' from December 2nd to December 21st, 2022}
  \label{fig:intensity_real}
\end{figure*}

\begin{table}[htbp]
\caption{Parameter Estimation: Real Twitter Dataset}
\begin{center}
\begin{tabular}{|c|c|c|}
\hline
\multicolumn{2}{|c|}{\textbf{Parameter Names}}&\textbf{Estimation} \\
\hline
\textbf{Immigrant Rate}& $\mu_s$&1.6917\\
\cline{2-3}
 $\boldsymbol{\mu_k}$& $\mu_n$ &0.3771\\
\hline
\textbf{Truncated}& lower bound&/\\
\cline{2-3} 
\textbf{Exponential}& upper bound&/ \\
\cline{2-3} 
\textbf{Distribution} & scale $x$&76.9602 \\
\hline
& $\delta_{ori}$&$1.5560\times10^{-3}$ \\
\cline{2-3} 
\textbf{Influence Factor}&$\delta_{ret}$&$4.1126\times10^{-3}$  \\
\cline{2-3} 
\textbf{of Tweet Type} & $\delta_{quo}$&$3.4947\times10^{-45}$  \\
\cline{2-3} 
$\boldsymbol{\delta_r}$& $\delta_{rply}$&$2.5179\times10^{-5}$ \\
\hline
& $\gamma_{ss}$&0.7763  \\
\cline{2-3} 
\textbf{Influence Factor}& $\gamma_{sn}$&0.2237  \\
\cline{2-3} 
\textbf{between Stances} & $\gamma_{ns}$&0.5853\\
\cline{2-3} 
$\boldsymbol{\gamma_{k',k}}$& $\gamma_{nn}$&0.4147\\
\hline
\textbf{Decay Parameter}&$\omega_s$&2.7141\\
\cline{2-3}
$\boldsymbol{\omega_k}$& $\omega_n$&4.8605\\
\hline
\end{tabular}
\label{tab:est_real}
\end{center}
\end{table}

Figure \ref{fig:intensity_real} presents the intensity of the dissemination process of the fake news story we collected. As we introduced in Section \ref{method_model_mul-hawkes}, we model the overall process by modeling the process of each specific stance considering the interplay between stances. The black curve shows the overall intensity, which is composed of the intensity of supporting and not-supporting stances recognized by red and green colors respectively. The red curve, corresponding to the estimated intensity of the supporting tweets, holds a higher intensity from the beginning to the end, which indicates that although the fact-checking articles are posted and disseminated starting from December 5th, tweets with a supporting stance toward the fake news story entirely dominated the dissemination process, where the fact-checking articles showed lower attention from the Twitter users.

Table \ref{tab:est_real} shows the estimation result of the parameters from modeling on the real dataset, which helps us understand how the user stances and tweet types influence the propagation of the fake news story quantitatively. The base immigrant rates for supporting and not-supporting stances are at the same level but with a higher value on $\mu_n$ which shows that original tweets with no obvious supporting stance are slightly more than the original tweets with obvious supporting stance. This indicates that there are an estimated 1.6917 supporting immigrants generated every 1 hour on average, and 0.3771 not-supporting immigrants generated every 1 hour on average which corresponds to 1 supporting original tweet being generated for every 35 minutes on average, and 1 not-supporting original tweet being generated for every 159 minutes on average.
The estimation result scale parameter $x$ for the truncated exponential distribution is about 75.9602, which corresponds to the meaning that the average arrival time of immigrants without stance-specified is about 76 hours.

The influence factor between stances $\gamma_{k',k}$ shows that a supporting tweet has an estimated $0.7763$ probability to trigger a supporting tweet and a $0.2237$ probability to trigger a not-supporting stance, while a not-supporting tweet has an estimated $0.4147$ probability of triggering a not-supporting tweet and a relatively higher probability ($0.5853$) of triggering a supporting tweet. This result is similar to what we assumed in the values determined for the parameters in the simulated dataset, which indicates that people tend to believe the fake news story in the current dissemination process of the fake news of covid vaccine will cause female sterilization, as both supporting and not-supporting stances hold a high probability of triggering a supporting tweet generated by a user that connects to the user who generated the prior supporting/not-supporting tweet. $\gamma_{ns}$ and $\gamma_{nn}$ are similar to our expectation that a not-supporting tweet will have about a half chance to generate a not-supporting tweet and a slightly more chance for generating a supporting tweet, which matches our observation of the tweets in the dataset where many users generated supporting quotes or replies after reviewing the fact-check.

The influence factor of each tweet type $\delta_r$ shows a similar trend but slightly deviates from what we expected from the simulated dataset. The original tweets take the dominant place in affecting users ($\delta_{ori} = 1.5560\times10^{-3}$) in terms of generating a tweet related to the fake news story, while retweets hold an even higher influence rate ($\delta_{ret} = 4.1126\times10^{-3}$) than the original tweets, by observing multiple retweets from a users' networks and become overwhelmed as the retweets take the smallest cost to be generated than quotes and replies, which correspond to the meaning that 1 tweet will be generated for every 643 users who observe an original tweet on average, and 1 tweet will be generated every 243 users who observe a retweet on average. This also demonstrates the low influence rates for quotes and replies that these two tweet types are generated with a higher cost as it requires comments from users. Replies also require users to click on the tweet and review them, and slide pages for more replies posted in the past such that users may easily lose their patience when replies are generated quickly or with a complex reply chain.

The decay parameters show the overall influence duration of each stance towards the fake news story. The supporting stance 
has a shorter impact ($\omega_s=2.7141$) than the not-supporting stance 
($\omega_n=4.8605$). 
This indicates that supporting tweets are triggered on average 22 minutes after the prior tweet that influenced it and non-supporting tweets occur 12 minutes after the triggering tweet. 

These estimated decay parameters indicate that tweets in a supporting stance hold a longer influence.

Taken all together these estimated parameters suggest that users tend to believe the fake news instead of the fact-check, or the fact-check articles posted several days later did not attract users' attention anymore. This result reveals the fact that fact-check articles make little effect on combating the fake news in the current fake news dissemination process, and matches the statement in \cite{ref30} that the correction from the fact-check websites cannot entirely revert public opinion to its original status.

\section{Discussion}\label{dis}
This paper derives the parameter estimation procedures and expressions for a novel Multivariate Hawkes Point Processes model for the fake news dissemination process using an Expectation Maximization approach. 

Our parameter estimation procedures were validated using a realistic simulated dataset. Subsequently, the procedures were applied to a real dataset of fake news collected from Twitter, demonstrating the practical application of the model in investigating the influence of user stances and tweet types during the dissemination process.

There are a few directions for possible future work. 
Because we collected data after the fake news event, our user network was not fully complete. Although we were able to infer many relationships from the collected tweets, the information is not perfect. Real-time data collection through using the Twitter streaming service could improve the quality of the user network information.  
Also, the text content of the tweets from users may hold different magnitudes on influencing users and triggering subsequent tweets. Quotes and replies posted by users will reveal their own opinions toward the fake news story from their comments, which can initiate connected users to respond to the comments. Therefore, the quantization of tweet text such as sentiment features can be considered to differentiate the ability of different tweets in affecting users and the dissemination process.
Furthermore, while we collected all the tweets related to the same fake news event, there were multiple variations and sub-topics within the fake-news story. 
Incorporating this additional information could help models better capture the true complexity and dynamics of the information flow. 

Our model is the first to explore the influence patterns between user stances during the fake news dissemination process, as well as quantifies the magnitude of the influence of different tweet types, which can be observed and illustrated by the real Twitter dataset.
Since the model can quantify how much each user was influenced by their user network to generate tweets taking each stance, it can be used to score each user as how likely they believe the fake news story. 
This likelihood of being persuaded by the misinformation/disinformation can be further investigated by exploring the users' past online behaviors such as propagating misinformation/disinformation or following unreliable users. 

{\appendices
\section{Maximum Likelihood Estimation}\label{app:mle}
Based on the model and the terms defined associated with different factors of fake news dissemination on Twitter, we can calculate the Likelihood Function for the intensity of events in stance $k$:
\begin{equation}
    L_k(t;\theta) = \prod_{j=1}^{N(t)}\lambda_k(t_j;\theta)e^{-\int_0^T\lambda_k(t;\theta)\mathrm{d}t}
\end{equation}
such that the log-likelihood function in stance $k$ can be expressed as follow:
\begin{align}
&\operatorname{log}L_{k}(t;\theta) = -\int_0^T\lambda_{k}(t;\theta)\mathrm{d}t+\sum_{j=1}^{N(t)}\operatorname{log}\lambda_{k}(t;\theta)\\
&\begin{aligned}[t]
=&-\int_0^T\left(\mu_k\frac{xe^{-xt}}{1-e^{-x}}+\sum_{j=1}^{N(t)}\delta_{r_j}\gamma_{k_j,k}n_jg_k(t-t_j)\right)\mathrm{d}t\\
&+\sum_{j=1}^{N(t)}\operatorname{log}\left(\mu_k\frac{xe^{-xt}}{1-e^{-x}}+\sum_{l=1}^{N(t_j)}\delta_{r_l}\gamma_{k_l,k_j}n_lg_k(t_j-t_l)\right)
\end{aligned}\\
&\begin{aligned}[t]
=&-\mu_k-\sum_{j=1}^{N(t)}\int_{t_j}^{T}\delta_{r_j}\gamma_{k_j,k}n_jg_k(t-t_j)\mathrm{d}t\\
&+\sum_{j=1}^{N(t)}\operatorname{log}\left(\mu_k\frac{xe^{-xt}}{1-e^{-xT}}+\sum_{l=1}^{j-1}\delta_{r_l}\gamma_{k_l,k_j}n_lg_k(t_j-t_l)\right)    
\end{aligned}
\end{align}
Therefore, the log-likelihood function for the overall intensity can be expressed as follow:
\begin{align}
&\operatorname{log}L(t;\theta) = \sum_{k\in K}\operatorname{log}L_{k}(t;\theta)\\
&\begin{aligned}
=&\sum_{k\in K}\Bigg[-\mu_k-\sum_{j=1}^{N(t)}\int_{t_j}^{T}\delta_{r_j}\gamma_{k_j,k}n_jg_k(t-t_j)\mathrm{d}t\\
&+\sum_{j=1}^{N(t)}\operatorname{log}\left(\mu_k\frac{xe^{-xt}}{1-e^{-xT}}+\sum_{l=1}^{j-1}\delta_{r_l}\gamma_{k_l,k_j}n_lg_k(t_j-t_l)\right)\Bigg]
\end{aligned}
\end{align}

\section{Expectation Maximization: $Q$ function}\label{app:q}
As aforementioned, E-step calculates the expectation of the log-likelihood function, which is also called the $Q$ function. Based on Jensen's inequality, we performed the following transformation for the log-likelihood function in stance $k$:
\begin{align}
&\operatorname{log}L_{k}(t;\theta)=\sum_{j=1}^{N(t)}\operatorname{log}\left(\mu_k(t)+\sum_{l=1}^{N(t_j)}h_{k}(t_j-t_l;\theta)\right)\\
&\quad -\int_0^T\left(\mu_k(t)+\sum_{j=1}^{N(t)}h_{k}(t-t_j;\theta)\right)\mathrm{d}t\\
&\begin{aligned}
=&\sum_{j=1}^{N(t)}\operatorname{log}\left(p_{jj}^{k}\cdot\frac{\mu_k(t)}{p_{jj}^{k}}+\sum_{l=1}^{N(t_j)}p_{jl}^{k}\cdot\frac{h_k(t_j-t_l;\theta)}{p_{jl}^{k}}\right)\\
&-\int_0^T\left(\mu_k(t)+\sum_{j=1}^{N(t)}h_{k}(t-t_j;\theta)\right)\mathrm{d}t
\end{aligned}\\
&\begin{aligned}
\geq&\sum_{j=1}^{N(t)}\left[p_{jj}^{k}\operatorname{log}\left(\frac{\mu_k(t)}{p_{jj}^{k}}\right)+\sum_{l=1}^{N(t_j)}p_{jl}^{k}\operatorname{log}\left(\frac{h_{k}(t_j-t_l;\theta)}{p_{jl}^{k}}\right)\right]\\
&-\int_0^T\left(\mu_k(t)+\sum_{j=1}^{N(t)}h_{k}(t-t_j;\theta)\right)\mathrm{d}t\\
\end{aligned}\\
&\begin{aligned}
=&\sum_{j=1}^{N(t)}p_{jj}^{k}\operatorname{log}(\mu_k\frac{xe^{-xt}}{1-e^{-xT}})-\sum_{j=1}^{N(t)}\int_{t_j}^Th_{k}(t-t_j;\theta)\mathrm{d}t\\
&+\sum_{j=1}^{N(t)}\sum_{l=1}^{N(t_j)}p_{jl}^{k}\operatorname{log} h_{k}(t_j-t_l;\theta)-\mu_k+C  \label{eq:Q_k}
\end{aligned}
\end{align}
where 
\begin{equation}
\sum_{j=1}^{N(t)}h_{k}(t-t_{j};\theta) = \sum_{j=1}^{N(t)}\delta_{r_j}\gamma_{k_j,k}n_jg_k(t-t_j)
\end{equation}
refers to the self-exciting function which models the intensity of each of the prior event (tweet) $j$ at time $t$ with respect to the pre-defined parameters.
Thus, $Q$ function in stance $k$ takes the form of equation \ref{eq:Q_k}:
\begin{equation}
\begin{split}
Q_k(T;\theta) =&\sum_{j=1}^{N(t)}p_{jj}^{k}\operatorname{log}(\mu_k\frac{xe^{-xt}}{1-e^{-xT}})-\mu_k\\
&+\sum_{j=1}^{N(t)}\sum_{l=1}^{N(t_j)}p_{jl}^{k}\delta_{r_j}\gamma_{k_j,k}n_jg_k(t-t_j)\\
&-\sum_{j=1}^{N(t)}\int_{t_j}^T\delta_{r_j}\gamma_{k_j,k}n_jg_k(t-t_j)\mathrm{d}t+C
\end{split}
\end{equation}
and the overall $Q$ function is expressed as:
\begin{equation}
\begin{split}
Q(T;\theta) =&\sum_{k\in K}\sum_{j=1}^{N(t)}p_{jj}^{k}\operatorname{log}(\mu_k\frac{xe^{-xt}}{1-e^{-xT}})-\sum_{k\in K}\mu_k\\
&+\sum_{k\in K}\sum_{j=1}^{N(t)}\sum_{l=1}^{N(t_j)}p_{jl}^{k}\delta_{r_j}\gamma_{k_j,k}n_jg_k(t-t_j)\\
&-\sum_{k\in K}\sum_{j=1}^{N(t)}\int_{t_j}^T\delta_{r_j}\gamma_{k_j,k}n_jg_k(t-t_j)\mathrm{d}t+C
\end{split}
\end{equation}

\section{Expectation Maximization: Parameter Expression}\label{app:param}
The maximum value of parameter $\theta$ occurs when $\frac{\partial Q(T;\theta)}{\partial\theta}=0$ such that for $\mu_k$ in $Q_k(T;\theta)$ we have:
\begin{equation}
    \frac{\partial Q_k(T;\theta)}{\partial\mu_k}=\sum_{j=1}^{N(t)}p_{jj}^{k}\frac{1}{\mu_k\cdot\frac{xe^{-xt}}{1-e^{-xT}}}\cdot\frac{xe^{-xt}}{1-e^{-xT}}-1=0
\end{equation}
\begin{equation}
    \frac{1}{\mu_k}\sum_{j=1}^{N(T)}p_{jj}^{k}=1
\end{equation}
\begin{equation}
    \mu_k=\sum_{j=1}^{N(t)}p_{jj}^{k}
\end{equation}
Particularly, the number of immigrants follows a Poisson distribution with a mean of overall immigrant rate which equals to $\mu_kT$, such that the equation of updating $\mu_k$ at each iteration $s+1$ of calculation is equivalent to:
\begin{equation}
    \mu_k^{(s+1)}=\frac{\sum_{j=1}^{N(t)}{p_{jj}^{k}}^{(s)}}{T}
\end{equation}

For the other parameters associated with the self-exciting process when applying the exponential kernel function, we have:
\begin{align}
    \frac{\partial\left(\sum_{j=1}^{N(t)}\sum_{l=1}^{N(t_j)}p_{jl}^{k}\operatorname{log}h_{k}(t_j-t_l;\theta)\right)}{\partial\theta}\nonumber\\
    -\frac{\partial\left(\sum_{j=1}^{N(t)}\int_{t_j}^Th_{k}(t-t_j;\theta)\mathrm{d}t\right)}{\partial\theta}=0
\end{align}

For $\gamma_{k',k}$, we have:
\begin{equation}
    \sum_{j=1}^{N(t)}\sum_{l:k_l=k'}^{N(t_j)}p_{jl}^k\frac{1}{\gamma_{k',k}}=\sum_{j=1}^{N(t)}\delta_{r_j}n_j\left(1-e^{-\omega_k(T-t_j)})\right)
\end{equation}
\begin{equation}
    \gamma_{k',k}=\frac{\sum_{j=1}^{N(t)}\sum_{l:k_l=k'}^{N(t_j)}p_{jl}^k}{\sum_{j=1}^{N(t)}\delta_{r_j}n_j\left(1-e^{-\omega_k(T-t_j)})\right)}
\end{equation}
Consider the case that a fake news story outbreaks on Twitter initially, diminishes over time, and vanishes eventually, the occurring time of each event $t_j$ should deviate from the total time $T$ of the observation as the intensity decays over time, such that $e^{-\omega_k(T-t_j)}\approx0$. Thus, the above expression can be simplified to its closed-form expression:
\begin{align}
    \gamma_{k',k}\approx\frac{\sum_{j=1}^{N(t)}\sum_{l:k_l=k'}^{N(t_j)}p_{jl}^k}{\sum_{j=1}^{N(t)}\delta_{r_j}n_j}
\end{align}
Particularly, as the assumption \ref{asp:ret} states, all the retweets should hold the same stance as the tweet triggers it, such that the between-stance parameter $\gamma_{k',k}$ should take effect on generating quotes and replies with the following adjustment on the expression at each iteration $s+1$:
\begin{align}
    {\gamma_{k',k}}^{(s+1)}=\frac{\sum_{j:r_j\in\{quo,rply\}}^{N(t)}\sum_{l:k_l=k'}^{j-1}{p_{jl}^k}^{(s)}}{\sum_{k\in K}\sum_{j:r_j\in\{quo,rply\}}^{N(t)}\sum_{l:k_l=k'}^{j-1}{p_{jl}^k}^{(s)}}
\end{align}

Similarly, for $\omega_k$, we have:
\begin{equation}
\begin{split}
    &\sum_{j=1}^{N(t)}\sum_{l=1}^{N(t_j)}p_{jl}^k\frac{\cdot\left(1-\omega_k(t_j-t_l)\right)}{\omega_k}=\\
    &\sum_{j=1}^{N(t)}\delta_{r_j}\gamma_{k_j,k}n_j(T-t_j)e^{-\omega_k(T-t_j)}
\end{split}
\end{equation}

\begin{align}
    &\omega_k=\\
    &\frac{\sum_{j=1}^{N(t)}\sum_{l=1}^{N(t_j)}p_{jl}^k}{\sum_{j=1}^{N(t)}\left[\sum_{l=1}^{N(t_j)}p_{jl}^k(t_j-t_l)+\delta_{r_j}\gamma_{k_j,k}n_j(T-t_j)e^{-\omega_k(T-t_j)}\right]}
\end{align}
such that we have the following simplified expression for $\omega_k$ to be updated at each iteration $s+1$:
\begin{align}
    \omega_k^{(s+1)}=\frac{\sum_{j=1}^{N(t)}\sum_{l=1}^{j-1}{p_{jl}^k}^{(s)}}{\sum_{j=1}^{N(t)}\sum_{l=1}^{j-1}{p_{jl}^k}^{(s)}(t_j-t_l)}
\end{align}

Since $\delta_r$ is associated with all user stance $k$ but specific to tweet type $r$ such that we need to derive its expression over the overall $Q(T;\theta)$ function:
\begin{equation}
    \sum_{k\in K}\sum_{j=1}^{N(t)}\sum_{l:r_l=r}^{N(t_j)}p_{jl}^k\frac{1}{\delta_r}=\sum_{k\in K}\sum_{j:r_j=r}^{N(t)}\gamma_{k_j,k}n_j\left(1-e^{-\omega_k(T-t_j)})\right)
\end{equation}
\begin{equation}
    \delta_r=\frac{\sum_{k\in K}\sum_{j=1}^{N(t)}\sum_{l:r_l=r}^{N(t_j)}p_{jl}^k}{\sum_{k\in K}\sum_{j:r_j=r}^{N(t)}\gamma_{k_j,k}n_j\left(1-e^{-\omega_k(T-t_j)})\right)}
\end{equation}
Hence, the closed-form solution for $\delta_r$ at each iteration $s+1$ is:
\begin{align}
    \delta_{r}^{(s+1)}=\frac{\sum_{k\in K}\sum_{j=1}^{N(t)}\sum_{l:r_l=r}^{j-1}{p_{jl}^k}^{(s)}}{\sum_{k\in K}\sum_{j:r_j=r}^{N(t)}\gamma_{k_j,k}^{(s)}n_j}
\end{align}
}

\vfill


\begin{thebibliography}{1}
\bibitem{ref25}
Ref25 should be the simulation paper which has not been published yet.
\end{thebibliography}
\end{document}